\documentclass[12pt,a4paper]{article}

\usepackage{amsmath}
\usepackage{amssymb}

\makeatletter
\let\old@makecaption=\@makecaption
\def\@makecaption{\small\old@makecaption}
\makeatother

\makeatletter
\let\old@startsection=\@startsection
\renewcommand{\@startsection}[6]{\old@startsection{#1}{#2}{#3}{#4}{#5}{#6\mathversion{bold}}}
\makeatother

\let\oldPhi=\Phi
\let\oldPsi=\Psi
\let\oldGamma=\Gamma
\let\oldSigma=\Sigma
\renewcommand{\Phi}{\mathnormal{\oldPhi}}
\renewcommand{\Psi}{\mathnormal{\oldPsi}}
\renewcommand{\Gamma}{\mathnormal{\oldGamma}}
\renewcommand{\Sigma}{\mathnormal{\oldSigma}}

\newcommand{\hypref}[2]{\ifx\href\asklfhas #2\else\href{#1}{#2}\fi}

\newcommand{\sfrac}[2]{{\textstyle\frac{#1}{#2}}}

\newcommand{\alg}[1]{\mathfrak{#1}}

\newcommand{\alSU}{\alg{su}}

\newcommand{\order}[1]{\mathcal{O}(#1)}

\newcommand{\superN}{\mathcal{N}}
\newcommand{\gym}{g_{\scriptscriptstyle\mathrm{YM}}}
\newcommand{\gnorm}{g}

\newcommand{\Tr}{\mathop{\mathrm{Tr}}}

\newcommand{\matr}[2]{\left(\begin{array}{#1}#2\end{array}\right)}

\newcommand{\lrbrk}[1]{\left(#1\right)}
\newcommand{\bigbrk}[1]{\bigl(#1\bigr)}

\newcommand{\comm}[2]{[#1,#2]}

\newcommand{\nln}{\nonumber\\}
\newcommand{\nl}{\nonumber\\&&\mathord{}}

\newcommand{\eq}{\mathrel{}&=&\mathrel{}}
\newenvironment{myeqnarray}{\arraycolsep0pt\begin{eqnarray}}{\end{eqnarray}\ignorespacesafterend}
\newenvironment{myeqnarray*}{\arraycolsep0pt\begin{eqnarray*}}{\end{eqnarray*}\ignorespacesafterend}

\def\[{\begin{equation}}
\def\]{\end{equation}}
\def\<{\begin{myeqnarray}}
\def\>{\end{myeqnarray}}

\begin{document}

\begin{flushright}\footnotesize
\texttt{hep-th/0308074}\\
\texttt{AEI 2003-070}
\end{flushright}
\vspace{0cm}

\begin{center}
{\Large\textbf{\mathversion{bold}
Higher loops, integrability\\and the near BMN limit}\par}
\vspace{1cm}

\textsc{Niklas Beisert}
\vspace{5mm}

\textit{Max-Planck-Institut f\"ur Gravitationsphysik\\
Albert-Einstein-Institut\\
Am M\"uhlenberg 1, D-14476 Golm, Germany}
\vspace{3mm}

\texttt{nbeisert@aei.mpg.de}\par\vspace{1cm}

\textbf{Abstract}\vspace{7mm}

\begin{minipage}{13.0cm}
In this note we consider higher-loop contributions to the planar 
dilatation operator of $\superN=4$ SYM in the $\alSU(2)$ subsector
of two complex scalar fields.
We investigate the constraints on the form of this object 
due to interactions of two excitations in the BMN limit.
We then consider two scenarios to uniquely fix some
higher-loop contributions:
(i) Higher-loop integrability fixes the dilatation generator 
up to at least four-loops. Among other results, this allows 
to conjecture an all-loop expression for the energy in the near BMN limit.
(ii) The near plane-wave limit of string theory and the BMN correspondence 
fix the dilatation generator up to three-loops.
We comment on the difference between both scenarios.
\end{minipage}

\end{center}
\vspace{1cm}
\hrule height 0.75pt
\vspace{1cm}

The complexity of perturbative calculations in quantum 
field theories increases dramatically with the 
loop order. While one-loop and perhaps two-loop 
Feynman diagrams can be worked out by hand, 
complete two-loop and three-loop calculations
are accessible by means of computer algebra systems.
Soon enough higher-loop investigations exceed the
capacities of today's and tomorrow's computers.
An immediate question is of course,
\emph{Do we need higher-loops?}
In some sense it does not really matter whether 
we know $n$-loops or $(n+1)$-loops. However, 
the first few terms in a sequence sometimes
enables one to guess how it continues.
Knowing, e.g., the first four numbers of the sequence instead of three,
reasonably increases the chances of a good guess.
In some cases it might be possible to recast the sequence 
as a function of the coupling constant.
Neglecting other, non-perturbative effects,
this function can then be investigated in the strong
coupling regime and might provide a test
of the (weak/strong) AdS/CFT correspondence.
\bigskip

In this note we follow the lines of
\cite{Beisert:2003tq} and investigate 
the planar dilatation generator of $\superN=4$ Super Yang-Mills
theory in the $\alSU(2)$ subsector consisting of 
two charged scalars $Z,\phi$. 
The philosophy of that work was to write down the most 
general operator allowed by the Feynman diagrams that give 
rise to anomalous dimensions. The coefficients were then
determined by fitting to known data. 
In the notation of that paper the two-loop dilatation operator 
was found to be
\<\label{eq:D024}
D(\gnorm)\eq \sum_{\ell=0}^\infty \lrbrk{\frac{\gym^2 N}{16\pi^2}}^\ell\,D_{2\ell},
\nln
D_0\eq \{\},
\nln
D_2\eq 2\{\}-2\{0\},
\nln
D_4\eq -8\{\}+12\{0\}-2\lrbrk{\{0,1\}+\{1,0\}},
\nln
\{n_1,n_2,\ldots\}\eq
\sum_{k=1}^L P_{k+n_1,k+n_1+1}
P_{k+n_2,k+n_2+1}\ldots
\>
where $P_{k,k+1}$ interchanges the scalars at two adjacent sites.

To determine the unknown coefficients, the BMN limit 
was used as input. We consider a state $\Tr Z^J \phi^k+\mbox{perm.}$ 
with $k$ excitations $\phi$ in 
a background of $J$ background fields $Z$, where $k\ll J$.
The BMN limit teaches us that the $\ell$-loop anomalous dimension 
$\delta\Delta_{2\ell}$ should scale as 
\cite{Berenstein:2002jq}
\[\label{eq:limit}
\delta\Delta_{2\ell}\sim (\lambda')^\ell\bigbrk{1+\order{1/J}},
\qquad 
\lambda'=\frac{\gym^2 N}{J^2}.
\]
To be more precise, also the coefficients are known from 
plane wave string theory \cite{Berenstein:2002jq}
\[\label{eq:energy}
\Delta=J+\sum_{i=1}^k \sqrt{1+\lambda'\,n_k^2}+\order{1/J},
\]
subject to the level matching constraint
$\sum_{i=1}^k n_k=0$. This all-loop conjecture was confirmed 
using gauge theory means in \cite{Santambrogio:2002sb}.

The limit \eqref{eq:limit} and the energy formula \eqref{eq:energy}
constrain the most general dilatation operator. It is easily seen that
in a dilute gas of excitations, 
the dilatation operator should act on the position of a single $\phi$
in a background of $Z$'s as \cite{Gross:2002su}
\[\label{eq:singleprop}
D_{2\ell}\approx -2 C_\ell\,\Box^\ell,
\]
where $\Box$ is the lattice Laplacian.
In the dilute gas approximation \eqref{eq:singleprop} gives
\[\label{eq:dilute}
\delta\Delta_{2\ell}=(-1)^{i+1}\, 2^{1-2\ell}C_\ell \sum_{i=1}^k \bigbrk{\lambda'\,n_k^2}^\ell
\]
which matches exactly \eqref{eq:energy}
when $C_{1,2,3,4,\ldots}=(1,1,2,5,\ldots)$ are the
Catalan numbers governing the expansion of the square root. 
Using this constraint sufficed to determine all
coefficients of the two-loop operator \eqref{eq:D024}. 

For the case of two excitations, $\Tr Z^J\phi^2+\mbox{perm.}$, the one-loop 
dilatation operator can be diagonalised explicitly \cite{Beisert:2002tn}. 
Once this is done, all higher-loop anomalous dimensions
can be obtained exactly in perturbation theory.
The all-loop generalisation of the three-loop energy formula presented
in \cite{Beisert:2003tq} appears to be
\footnote{Note that the expression $\Delta^{J}_n$ gives the scaling dimensions of the 
superconformal primary states with two excitations.
Here, $2+\Delta^{J-2}_n$ gives the scaling dimension
for the descendant state $\Tr Z^J\phi^2+\mbox{perm.}$.}
\[\label{eq:TwoExcitations}
\Delta^J_n:=J+2
+\sum_{\ell=1}^\infty 
\lrbrk{\frac{\gym^2 N}{\pi^2}\sin^2 \frac{\pi n}{J+3}}^{\ell}
\lrbrk{c_\ell+\sum_{k,m=1}^{\ell-1}c_{\ell,k,m}\frac{\cos^{2m} \frac{\pi n}{J+3}}{(J+3)^k}}
\]
where the first few coefficients derived from 
\eqref{eq:D024} are given by
\[
c_1=1,\qquad
c_2=-\sfrac{1}{4},\quad
c_{2,1,1}=-1.
\]
This formula interpolates smoothly between the regime 
of large $J$ all the way down to the Konishi operator.
Reassuringly, \eqref{eq:TwoExcitations} gives the correct
two-loop scaling dimension $\Delta_0^1$ of the Konishi
operator!

In order to fix the three-loop contribution, the BMN limit turned out 
to be not sufficient.
There are six free coefficients and \eqref{eq:singleprop} fixes
only four of them. 
Further insights were needed to find the remaining coefficients.
It was shown by Minahan and Zarembo 
that the planar dilatation generator
in the subsector of scalars, including the $\alSU(2)$ subsector under 
consideration, is integrable \cite{Minahan:2002ve}.%
\footnote{Recently, this result was extended to the full one-loop planar
$\superN=4$ SYM theory \cite{Beisert:2003yb}.}
Interestingly, this integrability seems to extend
(in a perturbative sense) also to the two-loop contribution
\eqref{eq:D024} \cite{Beisert:2003tq}. 
Therefore it is a reasonable assumption that
also the three-loop contribution would exhibit integrability.
This assumption was used to determine the remaining 
two coefficients of the three-loop dilatation operator.
It was attempted to go to four-loops in this way, unfortunately,
it seemed that a \emph{single} coefficient out of twelve could not be
determined.
\bigskip

However, this is not the case: As we will see, even the 
\emph{four-loop} contribution to the dilatation generator is uniquely 
fixed using (a) the BMN limit and (b) integrability!
The constraint \eqref{eq:singleprop} 
governing the propagation of a single excitation is a 
\emph{necessary} condition for the BMN limit, it is however,
not \emph{sufficient}.
There are further constraints which were overlooked in 
\cite{Beisert:2003tq}.
Namely, also the interactions of two or more excitation must obey a very specific
pattern in order not to spoil \eqref{eq:limit}.
This can be seen in the following way. 
For pairwise interactions the excitations must be close,
due to phase space considerations these interactions 
are then suppressed by $1/J$.
For the BMN-limit \eqref{eq:limit}, however,
they would need to be suppressed by $1/J^{2\ell+1}$.
Indeed, for $D_2,D_4$ this happens to be the case.
In contrast, for $D_6$ the most general contribution 
which obeys \eqref{eq:singleprop}, gives
$\delta\Delta_{6}\sim 1/J^5$ in violation of \eqref{eq:limit}.
Only if the coefficients of $D_6$ are arranged in a very specific way%
\footnote{It would be interesting to find a general criterion 
which determines whether an interaction
of two or more excitations respects the BMN limit or not.}
we will get $\delta\Delta_{6}\sim 1/J^6$.
This is fortunate, as we can impose novel constraints 
on the coefficients of the higher-loop contributions.
As we shall see, it allows to fix at least one further coefficient of $D_6$
from the BMN limit alone.
The same happens for the four-loop contribution;
in order to have $\delta\Delta_{8}\sim 1/J^8$ we can fix at least two 
additional coefficients.%
\medskip

Let us now reinvestigate the three-loop contribution $D_6$. 
We demand that the BMN limit \eqref{eq:limit}
exists, i.e. $\delta\Delta_6\sim 1/J^6$. Note that we will \emph{not} require
that the BMN energy formula \eqref{eq:energy} is reproduced correctly.
Four coefficients are fixed as follows
\<\label{eq:D6}
D_6\eq 
(60+6\alpha_1-56\alpha_2)\{\}
+(-104+14\alpha_1+96\alpha_2)\{0\}
\nl
+(24+2\alpha_1-24\alpha_2)\lrbrk{\{0,1\}+\{1,0\}}
+(4+6\alpha_1)\{0,2\}
\nl
+(-4+4\alpha_2)\lrbrk{\{0,1,2\}+\{2,1,0\}}
-\alpha_1\lrbrk{\{0,2,1\}+\{1,0,2\}}.
\>
From this we can derive the following coefficients for the energy formula 
\eqref{eq:TwoExcitations} of states with two excitations
\[
c_3=\sfrac{1}{8}-\sfrac{1}{8}\alpha_2,\quad
c_{3,k,m}=\matr{ll}{+\sfrac{3}{4}+\sfrac{3}{4}\alpha_1&+\sfrac{1}{2}-\alpha_1-2\alpha_2\\-\sfrac{3}{4}&+\sfrac{5}{2}}.
\]
At this point we need further input to fix the remaining coefficients.
There are two ways in which to proceed: (i) We rely on higher-loop integrability or,
(ii) we use the near plane-wave limit \cite{Parnachev:2002kk,Callan:2003xr}.%
\bigskip

In the case (i) we investigate degenerate pairs of operators \cite{Beisert:2003tq}.
For states of length $7,8$ with $3$ excitations there is
one degenerate pair each. Three-loop degeneracy requires
that
\[\alpha_1=\alpha_2=0\]
in agreement with the result of \cite{Beisert:2003tq}.
The other pairs are then also three-loop degenerate, which can be proved
by showing that the second integrable charge extends to three-loops
\cite{Beisert:2003tq}.
There are some interesting points regarding this solution.
For one, integrability fixes exactly the right number of 
coefficients for a unique solution. Moreover, the constraints from
integrability are compatible with the constraints from the BMN limit.
Most importantly, we have only demanded the \emph{qualitative}
BMN limit \eqref{eq:limit}. Integrability fixes the remaining coefficients
in just the right way for the \emph{quantitative} 
BMN limit \eqref{eq:energy}, i.e. $c_3=+\sfrac{1}{8}$.
\medskip

As the procedure seemed very successful at the three-loop level 
we now turn to four-loops. There we have $12$ distinct structures
and corresponding coefficients. Here, the existence of the BMN
limit determines $6$ coefficients, $4$ from the propagation of a
single excitation and $2$ from pairwise interactions. 
Integrability determines $5$ further coefficients.
As explained in \cite{Beisert:2003tq}, 
the remaining coefficient $\beta$ multiplies the structure
$\comm{\comm{D_4}{D_2}}{D_2}$; it corresponds to a rotation
of the space of states generated by $\gnorm^6\comm{D_4}{D_2}$
and thus does not influence scaling dimensions.
The resulting four-loop dilatation generator is
\<
D_8\eq
-560  \{\}
+(1036+4\beta)\{0\}
\nl
+(-266-4\beta) (\{0, 1\} + \{1, 0\})  
+(-66-2\beta)\{0, 2\}
-4 \{0, 3\}
\nl
+4(\{0, 1, 3\} + \{0, 2, 3\} + \{0, 3, 2\} + \{1, 0, 3\}) 
\nl
+(78+2\beta) (\{0, 1, 2\} + \{2, 1, 0\}) 
+(-18+2\beta) (\{0, 2, 1\} + \{1, 0, 2\})  
\nl
+(1-\beta) ( \{0, 1, 3, 2\} + \{0, 3, 2, 1\} + \{1, 0, 2, 3\} + \{2, 1, 0, 3\} )  
\nl
+(6-2\beta) \{1, 0, 2, 1\}
+2\beta (\{0, 2, 1, 3\} + \{1, 0, 3, 2\})  
\nl
-10(\{0, 1, 2, 3\} + \{3, 2, 1, 0\}),
\>
it equals the structure given in \cite{Beisert:2003tq} with $\alpha=3$.
The four-loop coefficients of the energy formula \eqref{eq:TwoExcitations} are
given by%
\footnote{Using these coefficients
the energy of the Konishi operator appears to be
\[\nonumber
\Delta_{\mathcal{K}}=
\Delta_{0}^1=
2+\frac{3\gym^2N}{4\pi^2}
-\frac{3\gym^4N^2}{16\pi^4}
+\frac{21\gym^6N^3}{256\pi^6}
-\frac{705\gym^8N^4}{16384\pi^8}+\ldots
\]
However, in this case it is not clear whether \eqref{eq:TwoExcitations}
applies, because the length of the interaction $D_8$, $5$,
exceed the length of the state, $4$. The above expression 
therefore represents an extrapolation of 
the range of validity of $D_8$, which might or might not be true.}
\[\label{eq:Coeff4}
c_4=-\sfrac{5}{64},\quad
c_{4,k,m}=\matr{lll}{
-\sfrac{5}{8}&-\sfrac{5}{12}&-\sfrac{1}{3}\\
+\sfrac{3}{4}&- \sfrac{7}{4}&-\sfrac{7}{2}\\
-\sfrac{1}{2}&+\sfrac{59}{12}&-\sfrac{49}{6}}.
\]
Again, we see that the BMN energy, $c_4=-\sfrac{5}{64}$, is
predicted correctly by integrability. Again, the BMN limit and
integrability fix a complementary set of coefficients. 
It remains to be seen whether also the five-loop dilatation generator 
can be obtained in this way; it involves approximately sixty structures.
\medskip

A further application of \eqref{eq:TwoExcitations} is the 
near BMN limit of $\order{1/J}$ corrections to the energy. 
Some inspired guessing yields an all-loop expression for the near BMN limit
which agrees at four-loops
\[\label{eq:nearBMN}
\Delta^J_n=J+2\sqrt{1+\lambda'\, n^2}
-\frac{8\lambda'\,n^2}{J\sqrt{1+\lambda'\,n^2}}
+\frac{2\lambda'\,n^2}{J(1+\lambda'\,n^2)}
+\order{1/J^2}.
\]
The first $1/J$ term can be regarded as a renormalisation of the coupling
constant. If we replace $J$ in the definition of $\lambda'$ by
$J+4$ this term can be absorbed into the leading order energy.%
\bigskip

Unfortunately, the formula \eqref{eq:nearBMN} does not agree with the 
expression for the near plane-wave limit derived in \cite{Callan:2003xr}
\footnote{The level splitting is determined by
the strict BMN limit, we will consider only the primary state.}
\[\label{eq:nearPW}
\Delta^J_n=J+
2\sqrt{1+\lambda'\, n^2}
-\frac{4\lambda'\,n^2}{J\sqrt{1+\lambda'\,n^2}}
-\frac{2\lambda'\,n^2}{J}
+\order{1/J^2}.
\]
It agrees with \eqref{eq:TwoExcitations} only up to two-loops \cite{Callan:2003xr}.
In scenario (ii) we achieve agreement
by matching the remaining coefficients $\alpha_1,\alpha_2$ 
at three-loops to \eqref{eq:nearPW} and get
\[
\alpha_1=2,\quad\alpha_2=0.
\]
This choice of coefficients 
lifts the degeneracy of pairs at the three-loop level
and thus violates integrability.
Furthermore, the input from the near plane-wave limit is not sufficient to
determine four-loops unambiguously.
\bigskip

We conclude that higher-loop integrability 
contradicts with the result \eqref{eq:nearPW} 
of \cite{Callan:2003xr}.
At this point it is not clear which information to use
in order to construct the three-loop dilatation generator $D_6$.
On the one hand, given the long list of verifications 
of the BMN correspondence, it is hard not to believe in scenario (ii).
Nevertheless, the all-loop conjecture \eqref{eq:nearBMN} of (i)
is rather similar to the near plane-wave result \eqref{eq:nearPW} of (ii). 
Maybe a slight modification of the relations is required
to match both expressions? E.g. one might imagine that
the effective coupling constant $\lambda'$ is defined using 
$J+2$ instead of $J$, which apparently makes one term agree.
On the other hand, a failure of higher-loop integrability 
would be disappointing as well.
Here, it was demonstrated that integrability and the existence of a BMN limit
go hand in hand to determine the dilatation generator up to four-loops.
What is more, they predict the leading order BMN energy correctly. 
This appears to be more than a coincidence and makes scenario (i) 
very attractive:
Whether or not related to $\superN=4$ SYM, this scenario requires further 
investigations.
Eventually, only a derivation of the three-loop dilatation generator 
with as few assumption as possible may decide in either direction.
As was demonstrated in \cite{Beisert:2003jj} the superconformal algebra 
imposes some valuable constraints that might pave the way.

\subsection*{Acknowledgments}

The author is grateful to T.~Klose, J.~H.~Schwarz, D.~Serban and M.~Staudacher
for useful discussions.
Ich danke der \emph{Studienstiftung des
deutschen Volkes} f\"ur die Unterst\"utzung durch ein 
Promotions\-f\"orderungsstipendium.






\end{document}